\documentstyle[aaspp4]{article}
\begin{document}
 
\lefthead{Elmegreen}
\righthead{Initial mass function in hierarchical clouds}
\slugcomment{scheduled for ApJ, vol. 515, April 10, 1999}
 
\title{The initial stellar mass function from random sampling
in hierarchical clouds II: statistical fluctuations and a mass dependence for
starbirth positions and times}

\author{ Bruce G. Elmegreen\altaffilmark{1}}
\altaffiltext{1}{IBM Research Division, T.J. Watson Research Center,
P.O. Box 218, Yorktown Heights, NY 10598, USA, bge@watson.ibm.com}

\begin{abstract}
 
Observed variations in the slope of the initial stellar mass function
(IMF) are shown to be consistent with a model, introduced previously, in
which the protostellar gas is randomly sampled from clouds with
self-similar hierarchical structure. RMS variations in the IMF slope
around the Salpeter value are $\pm0.4$ 
when only 100 stars are observed, and $\pm0.1$ when 1000 stars are
observed. Similar variations should be present in other stochastic models too.

The hierarchical-sampling model reproduces the tendency for
massive stars to form closer to the center of a cloud, at a time
somewhat later than the formation time of the lower mass stars. The
systematic variation in birth position results from the tendency for the
trunk and larger branches of the hierarchical tree of cloud structure
to lie closer to the
cloud center, while the variations in birth order result from the relative
infrequency of stars with larger masses. 

The hierarchical cloud sampling model has now reproduced most of the
reliably observed features of the cluster IMF. The power law part of the
IMF comes
from cloud hierarchical structure that is sampled during various star
formation processes with a relative rate proportional to the square root
of the local density. These processes include turbulence compression,
magnetic diffusion, gravitational collapse, and clump or wavepacket
coalescence, all of which have about this rate dependence. The low mass
flattening comes from the inability of gas to form stars below the
thermal Jeans mass at typical temperatures and pressures. The thermal
Jeans mass is the only relevant scale in the problem. Considerations of
heating and cooling processes indicate why the thermal Jeans mass should
be nearly constant in normal environments, and why this mass might
increase in starburst regions. In particular, the relative abundance of
high mass stars should increase where the average density of the
interstellar medium is very large; accompanying this increase should be
an increase in the average total efficiency of star formation.
Alternative models in which the rate of star formation is independent of
density and the local efficiency decreases systematically with
increasing stellar mass can also reproduce the IMF, but this is an
adjustable result, not a fundamental property of hierarchical cloud
structure, as is the preferred model.

The steep IMF in the extreme field is not explained by the 
model but other origins are suggested, including one in which massive
stars in low pressure environments
halt other star formation in their clouds. In this case, the
slope of the extreme field IMF is independent of the
slope of each component cluster IMF, and is given by 
$(\gamma-1)/\alpha$
for cloud mass function slope $-\gamma\sim-2$ and 
power law relation, $M_L\propto M_c^\alpha$,
between the largest star in a low-pressure cloud,
$M_L$, and the cloud mass, $M_c$.
A value of $\alpha\sim1/4$ is
required to explain the extreme field IMF
as a superposition of individual cluster IMFs. 
We note that the similarity between cluster IMFs and the average
IMF from global studies of galaxies implies that most stars
form in clusters and that massive stars
do not generally halt star formation
in the same cloud.

\end{abstract}

Subject headings: stars: formation -- stars: mass function -- 
ISM: structure 

\section{Introduction}

The initial stellar mass function (IMF) is approximately a power law for
intermediate (Salpeter 1955) to large masses (see review in Massey
1998), but it flattens at low mass (Miller \& Scalo 1979; Kroupa, Tout
\& Gilmore 1990; see review in Reid 1998), often down to the limit of
detection, which is sometimes as low as 0.1$M_\odot$ or lower (see
reviews in Lada, Lada \& Muench 1998; Scalo 1986, 1998). Such flattening
is widely observed for Milky Way disk stars, and it is also inferred for
halo, bulge, and globular cluster stars (see reviews by Cool 1998 and
Wyse 1998). 

This flattened power-law IMF applies to a wide variety of regions,
although the details vary. For example, the mass at the threshold of the
flat part seems to be higher in starburst regions (Rieke et al. 1980,
1993; see reviews in Scalo 1990a; Zinnecker 1996; Leitherer 1998), giving
a larger proportion of high mass stars compared to the Solar
neighborhood. The cluster in 30 Doradus may have a higher threshold mass
too (Nota 1998). The result for starburst galactic nuclei comes from the
observation that the luminous mass from young massive stars is not much
smaller than the dynamical mass from the rotation curve, so there cannot
be much mass in the form of low mass stars (Wright et al. 1988; Telesco
1988; Doane \& Matthews 1993; Doyon, Joseph, \& Wright 1994; Smith et
al. 1995). However, Satyapal (1995,1997) found that the high mass stars
in the classical starburst galaxy M82 are spread out over a large area,
in which case the normal proportion of low mass stars could be
present. Other evidence for normal IMFs in starburst regions is in
Devereux (1989),
Schaerer (1996) and Calzetti (1997). The result for 30 Dor comes from
stellar counting, and if confirmed, would be the first direct evidence
for this effect.

There is also an observation that the largest clouds produce the largest
stars, as if the IMF depends on cloud mass (Larson 1982). This result is
analogous to the statement that spiral arm clouds, which tend to be
large, produce proportionally more massive stars than interarm clouds
(Mezger \& Smith 1977) which tend to be small. These concepts, along
with massive-star triggering scenarios, led to suggestions of a bi-modal
IMF (Eggen 1976, 1977; Elmegreen 1978; G\"usten \& Mezger 1983; Larson 1977, 1986).

The proportion of high and low mass stars may differ for cluster and
field populations too, in the sense that the slope of the local 
field star IMF
may be steeper than the slope of the cluster IMF (see discussion in
Elmegreen 1997, hereafter Paper I, and 1998). This is the case for
intermediate mass stars in the solar neighborhood (Scalo 1987)
and in an extreme field region of the LMC (Massey et al.
1995).

The most massive stars might also be born slightly closer to a cloud's
center than the low mass stars. In the Orion trapezium region,
Hillenbrand \& Hartmann (1998) found such a distribution and suggested
the stars must have been born this way because the cluster is too young
to have relaxed after an initially random birth order (see also Jones \&
Walker 1988; Hillenbrand 1997; Bonnell \& Davies 1998). High mass stars
are also concentrated in NGC 2157 (Fischer et al. 1998), SL 666, and NGC
2098 (Kontizas et al. 1998). This relative concentration of high mass
stars in clusters has been suspected for a long time on the basis of
color and IMF variations (Sagar \& Bhatt 1989; Pandey, Mahra, \& Sagar
1992; Vazquez et al. 1996). In addition, the massive stars in a cluster
might be born later than the low mass stars (Herbig 1962; Iben \& Talbot
1966), as recently suggested for 30 Dor (Massey \& Hunter 1998). 

These systematic variations in the IMF are distinct from the seemingly random
fluctuations from cluster to cluster. In a recent review, Scalo (1998)
pointed out that while the average slope of the IMF at intermediate to
high mass is about the value found originally by Salpeter, i.e., $-1.35$
for stellar counting in equal logarithmic intervals, the slopes for
individual regions vary by $\pm0.5$. It is currently unknown if these
variations result from physical differences in the intrinsic IMFs for 
each region, or from statistical fluctuations around a universal IMF.  
The extreme cases certainly look non-random since they have
smooth power law IMFs with slopes that are significantly different from 
the Salpeter value. For example, Brown (1998) found an IMF slope
in the Orion association of $\sim-1.8$, using Hipparcos data 
to determine membership combined with
photometry to determine masses, while Massey (1998) and collaborators 
get a fairly consistent value of $\sim-1.35$ for 
high mass stars in most associations in 
the Large Magellanic Cloud and the Milky Way, using spectroscopy
to determine masses.  The difference between these two results 
might be from the different methods used to determine mass (Massey 1998), 
or they could be from different physical processes of star formation.  
Another extreme example is the Pleiades IMF (Meusinger et al. 1996),
with its slope of $\sim-2.3$ (see discussion in Scalo 1998). Such 
steep slopes for clusters (see also NGC 6231 -- Sung, Bessel, \& Lee 1998)
are troublesome because they could arise from mass segregation and not
IMF differences (e.g., see Pandey, Mahra, \& Sagar 1992).

The purpose of the present paper is to consider 
some of these variations using
the model introduced in Paper I. We contend that not all of the
variations result from physical differences: some probably reflect
sampling limitations in the observations, others may result from
incorrect assumptions in the interpretation, and still others might be
purely statistical in nature. Yet some variations seem real and we would
like to explain them, as well as the statistical effects, with a minimum
of assumptions. 

Other theories of the IMF were reviewed by Cayrel (1990) and Clarke
(1998). A review of both observations and theory is in Elmegreen (1999).
The closest models to the present were by Henriksen (1986, 1991) and
Larson (1992). Henriksen used the size distribution function for
structures in a fractal of dimension $D$, given as $n(R)d\log R\propto
R^{-D}d\log R$ by Mandelbrot (1983), and assumed that the density $\rho$
varies with size as observed in self-gravitating clouds, $\rho\propto
R^{-1}$. This gave a mass function for cloud structure that agreed well
with observations if $D\sim2.7$. Larson (1992) also used the size
distribution for a fractal, and assumed that final stellar mass is
proportional to the linear size of the cloud piece.
Then the slope of the IMF in linear intervals becomes
$1+x=1+D\sim3.3$ for $D=2.3$. This IMF is steeper than the
Salpeter IMF, which has $1+x=2.3$, but Larson suggested that perhaps stars
form in subparts of clouds where $D$ is smaller than 2.3, or that
larger structures have larger temperatures, which would break the
assumed linear relationship between star mass and scale size.

The present model is similar to these in the suggestion that the power
law part of the IMF comes from cloud geometry.  It is similar to
Larson's model also  because of the assumption that the lower mass limit is
related to the thermal Jean's mass. However, the present model differs from
the others in several important respects. First, the star mass is
assumed to be proportional to the cloud mass here, rather than linear size, 
so we are not constrained to assume filamentary cloud pieces.
Second, the density structure of the cloud is given by the
hierarchy, consistent with the number-size relation, not by the density
structure 
of a self-gravitating cloud. As a result, the IMF
in the present model 
has a power law that depends only weakly on the fractal dimension
of the cloud, unlike the other models, and yet agrees with the observed
IMF in spite of a nearly complete absence of free parameters. 

We return to a discussion of the present model in Section
\ref{sect:theory}. In the next section, the various observations of the
IMF are reviewed in more detail in order to assess which are likely to
be the result of physical effects, and which are likely to come from
statistical sampling.

\section{Sorting out what's real}

The observation that high mass clouds produce high mass
stars is the type of effect that can arise from sampling statistics:
larger clouds produce more stars, so the IMF is sampled further out in
the high mass tail, giving a more massive, highest-mass star (Larson 1982). 
Indeed,
the statistical trend that is expected from random sampling matches the
observations well (Elmegreen 1983), so there is little room for further
effects based on physical variations (although see Khersonsky 1997).
Similarly, the inference that spiral arm clouds produce proportionally
more massive stars than interarm clouds should contain the same
sampling effect based on cloud mass, but in addition, could contain an
observational problem that low mass stars are too faint to see in
distant spiral arm clouds. 

Along these same lines, we also expect the most massive stars to form
last in a cluster because they are rare and unlikely to
appear until after several thousand lower mass stars form first
(Elmegreen 1983).

The difficulty in observing low mass stars may account for some of
the inference that starburst regions have truncated IMFs, but if the
truncation is severe and based on either direct star counts above the
sensitivity limits, or on luminous
mass compared to dynamical mass, then the argument is difficult to
interpret any other way.

The difference between field star IMFs and cluster IMFs is more
difficult to assess because both data are often of comparable quality,
and because there are three distinct types of observations that have led
to this conclusion. One is a comparison between the Solar neighborhood
field-star IMF (Scalo 1986; see also Tsujimoto et al. 1997 for
a discussion of metallicity yields and the field star IMF) 
or the LMC field star IMF (Parker et al. 1998), and essentially all
other IMFs measured in associations (Massey 1998). Another is a comparison
between IMFs of clustered star-forming regions in low and high density
environments (J.K. Hill et al. 1994; R.S. Hill et al. 1995; Massey et
al. 1995). The third is a comparison between high and low mass young
star counts in high and low density regions of the same cloud (Ali \&
DePoy 1995).

In the first case, there are corrections to the Solar field population
at low and intermediate mass that do not arise in young clusters --
corrections to account for stellar evolution, such as the loss of
evolved stars and possible variations in the past star formation rate
(Scalo 1986). There are also corrections to account for the systematic
increase in stellar scale height with age (Scalo 1986). These same
corrections arise at high mass too, but the time scales for evolution
and star formation variations are much smaller for high mass stars, and
the scale height effect could be erratic because of the small number of
stars involved. For example, at intermediate to high mass, one might
conjecture that the field star IMF is a composite of several different
cluster IMFs from aging dispersed associations in the Solar
neighborhood. Perhaps all of these local stars are from Gould's Belt,
with different ages, degrees of dispersal, and initial distances from
the plane for each former association. Then the Solar field star IMF at
intermediate and/or high mass could be steeper than the IMF in each
cluster if the low mass stars systematically drift further from their
points of origin than higher mass stars (because of the greater ages of
low mass stars), or if older clusters that have no massive stars anymore
(e.g., the Cas-Tau association) happen to be closer to us than younger
clusters that still have most of these stars (e.g., Perseus, Sco-Cen and
Orion). 

In the second case, there could be some effect from differential drift
as well, although Massey et al. (1995) contend that the extreme-field
regions in the LMC, in which star formation occurs 
more than 30 pc from cataloged
OB association boundaries, are too far from other clusters for low mass
contamination by differential drift to be important in the IMF. This is
puzzling since Massey \& Hunter (1998) found essentially the same IMF in
clusters with a wide range of densities. Thus, either the IMF is
independent of star-forming density, and the extreme field is somehow
not a representative sample, or there is a threshold low density where
the IMF abruptly changes from Salpeter-like to something much steeper at
lower density.

There is probably no evolutionary effect to explain the third case,
because the entire cloud is young, so differential drift does not seem
possible unless low mass stars move faster than high mass stars. On the
other hand, one might expect different IMFs in low and high density
regions of molecular clouds for the same reason that massive stars form
closer to the center of a cluster than low mass stars (Hillenbrand \&
Hartmann 1998). 

The last type of IMF variation discussed above, i.e., the scatter in the
slope of the IMF from cluster to cluster, is again something that might
be expected from statistical sampling. This is amenable to test because
one can look for a correlation between the numbers of stars used to
determine each IMF and the deviations of the slopes from the average
value for all clusters. If the slope deviation systematically decreases
for increasing numbers of stars, then the slope fluctuations could be
random. Otherwise, there is good reason to look for systematic physical
differences that give rise to the different slopes. 

We interpret these observations to indicate that, at the present
time, there are four distinct physical effects that have to be explained
by a theory of the IMF: (1) the power-law slope at intermediate to
high stellar mass, with a value close to that found by Salpeter (1955);
(2) the flattening at low mass for clusters and star-forming regions in
our Galaxy; (3) the increase in the transition mass between power-law
and flat regions of the IMF in starburst galaxies and 30 Dor, and (4) the
preferential birth of high mass stars close to cluster centers. We
also expect the theory to produce three additional effects for
statistical reasons: (1) the tendency for the largest mass star in a
region to increase with the total number of stars, (2) the tendency for
the largest mass stars to form last, and (3) the seemingly random
variations in the intermediate and high mass IMF slopes from region to
region. 

The first three of these physical effects, and the first of the
statistical effects, were demonstrated for our model in Paper I. They
will be discussed here again briefly in Section 2. The fourth physical
effect and the third statistical effect were not known at the time
Paper I was written, so they will be discussed here. The second
statistical effect has received recent attention by Massey \&
Hunter (1998), so it will be discussed here as well. We show that the
same IMF model is consistent with all three statistical and 
four physical effects, but the observation of the fourth physical 
effect is not matched well by the model.  There is some need
for an additional physical process, such as a reasonable enhancement 
in cloud condensation from self-gravity. This would alter the
internal structure of the cloud from a purely self-similar hierarchical cloud, 
as assumed by the model so far, to a centrally condensed structure. 
Such an alteration would not affect the
other attributes of the IMF model, including the {\it relative order} of
birth locations of stars in a cluster, but it would affect the {\it absolute}
birth locations as the cloud structure becomes condensed in the
center.

There could also be a physical effect that limits the masses of the
largest stars that form, but such a mass limit has apparently not been
observed yet (e.g., Massey \& Hunter 1998), so we do not consider it in
the model. Demonstration of a mass limit like this would require the
observation of an IMF out to very high masses, with a statistically
significant number of ``highest'' mass stars and an abrupt drop in the
number of stars with higher masses.

\section{The extreme field IMF}

The IMF model developed here and in Paper I 
cannot explain the steep slope of the high-mass IMF observed in
some extreme field studies (Massey et al. 1995) without considering
significantly different physical effects. Massey et al.
define the extreme field as regions of star formation
more than 30 pc from cataloged
OB associations. The extreme field regions of the Milky Way 
are not the same as the local field, which contains a mixture of
stars that were likely to have formed in standard clusters. 

Two possible explanations for the steep IMF in the extreme
field come to
mind. One is a systematic decrease in the efficiency of star formation
for increasing clump mass in the case of extreme field star formation,
but not clustered star formation. This is considered again in Section
\ref{sect:alpha}. This situation might arise if there is a transition at
extremely low pressure, such that in low pressure environments, all
stars destroy their clouds with increasing effectiveness at increasing
mass, but in high pressure environments, stars do not destroy their
clouds much at all, unless they are stellar types O or B. 

As an example of this effect, consider a field IMF that is a
superposition of IMFs from many star-forming clouds, all with different
masses, and consider that the largest stellar mass in each cloud
increases with the cloud mass, perhaps because it takes a more massive
star to destroy a more massive cloud. If $M_L$ is the largest stellar
mass and $M_c$ is the cloud mass, then we can write this condition as
$M_L\propto M_c^\alpha$. Because many clouds, even small clouds, can
produce low mass stars, but only the larger clouds produce high mass
stars, the summed IMF will be steeper than the individual cloud
IMFs. If the individual cloud IMF is $n(M)dM=n_0M^{-1-x}dM$,
then $n_0=xM_L^x$ so there is one star with the largest mass (i.e.,
$\int_{M_L}^\infty n(M)dM=1$). The summed IMF from all the clouds is
$n_{sum}(M)dM=n(M)\int_M^\infty P(M_L)dM_L$, where $P(M_L)dM_L$ is the
probability distribution for the largest mass, which is given by
$P(M_L)dM_L=N(M_c)dM_c$ for cloud mass distribution $N(M_c)\propto
M_c^{-\gamma}$. With the above 
relation between $M_L$ and $M_c$, $P(M_L)$ gets
converted to $P(M_L)dM_L\propto M_c^{-\gamma}\left(dM_c/dM_L\right)dM_L$,
which gives $P(M_L)\propto M_L^{(1-\gamma-\alpha)/\alpha}$. Thus the
integral for $n_{sum}$ can be evaluated, with the result that
$n_{sum}dM\propto M^{-1-x_{eff}}dM$ for effective
$x_{eff}=\left(\gamma-1\right)/\alpha$. Note that the summed IMF from
all the clusters is independent of the IMF of each cluster (i.e.,
$x_{eff}$ is independent of $x$). 

What are $\gamma$ and $\alpha$?
The power in the cloud mass function is $\gamma\sim2$, considering all
possible levels in the cloud hierarchy, as before, and the power in the
$M_L(M_c)$ relation presumably comes from cloud destruction. According to the
molecular cloud scaling relationships (Larson 1981), the binding energy
of a molecular cloud, $GM_c^2/R$, is proportional to radius $R^3$ since
$M_c\propto R^2$. Presumably an embedded
 cluster has to liberate a cloud binding
energy in a multiple of the crossing time in order to destroy it without
significant energy dissipation. Thus the cloud
destructive power of the embedded cluster has to
exceed $\left(GM_c^2/R\right)\left(GM_c/R^3\right)^{1/2}\propto
M_c^{5/4}$. This result assumes that low-pressure environments produce clouds
with the same scaling relations as normal star-forming
environments. To get the extreme field IMF, we need $\alpha\sim0.25$, so 
the cluster destructive power has to scale
with the fifth power of the largest stellar mass. 
Then the summed IMF has
$x_{eff}\sim\left(\gamma-1\right)/\alpha\sim4$, which is about what Massey et
al. (1995) see in the field regions of the LMC and Milky Way.
There are no observations of $\alpha$ in the extreme field, 
so this explanation requires confirmation.  The steep
slope of the relation between cluster destructive power 
and largest stellar mass 
might arise for clusters whose largest star is at the transition between A-type
and B-type; this is where the primary agent for cloud destruction changes
from pre-main sequence winds to ionization.

Note that in clustered regions, the IMF is about the same as in the
Solar neighborhood field, and also the same as the 
summed IMF in whole galaxies, i.e., all have $x\sim1.3-1.8$. This means
not only that most stars form in clusters and not the extreme field,
but also that {\it the most massive
stars do not halt star formation in their clouds by destroying the
clouds}. If they did, then the summed IMF would be steeper than the
individual cluster IMFs (for further discussion of this point, see the
review in Elmegreen 1999).

Another physical difference that could affect the IMF in the field
would be a difference in the type of star-forming cloud in low and high
density environments. Presumably low-density star-forming regions have
only small star-forming clouds, like the Taurus clouds, or only
low-density clouds, like the Maddalena \& Thaddeus (1985) cloud. Such
clouds may be systematically different than high-mass, high-density
clouds, such as those found in OB associations. It is possible, for
example, that low self-gravity and low central pressure allow low-mass
or low-density clouds to be more filamentary than high-mass or high
density clouds. In that case, gravitational instabilities along the
filaments can produce a {\it characteristic} mass for dense globules
(Tomisaka 1996), and stars forming in these globules (Nakamura et al.
1995) could produce a
delta-function IMF centered at some fraction of the globule mass. Such
an IMF would be very steep for high mass stars.

\section{Introduction to the theory}
\label{sect:theory}

A theory for the IMF based on random sampling in hierarchically
structured clouds has already been shown to have several of the
requirements discussed above (Paper I). The Salpeter slope comes from
two assumptions: (1) that star mass follows gas mass, and (2) gas turns
into stars at a rate proportional to some modest power of the local gas
density, such as a square root. This would be the case for
gravitational, magnetic diffusion-regulated, or turbulent processes, as
well as for clump-collision processes, considering the molecular cloud
scaling laws (see appendix). The low mass flattening comes from a third
assumption: (3) that gas clumps smaller than the thermal Jeans mass,
$M_J\sim (k_BT/m_{H2})^2/(G^3P)^{1/2}$, at the temperature $T$ and
pressure $P$ of the star-forming part of the cloud do not readily form
stars because they cannot ever become strongly self-gravitating.

Assumption (3) was claimed to be the reason why the transition mass
between the flat and power-law parts increases for starburst regions.
There the quantity $T^2/P^{1/2}$ presumably increases because of an
increase in $T$ with the higher radiation fields (offsetting the known
increase in $P$ for these regions). We also showed in Paper I, without
any other assumptions, how stochastic effects cause the mass of the
largest star to increase with the number of stars formed, which is the
first statistical effect mentioned above. Here we show that the theory
reproduces the observed level of fluctuations in the IMF slope from
region to region, as compiled by Scalo (1998), and we reproduce
the observation that
high mass stars form closer to a cluster center than low mass clouds
(Hillenbrand \& Hartmann 1998). We also show that high mass stars tend
to form last in a stochastic model like this.

The ratio $T^2/P^{1/2}$ that appears in the thermal Jeans' mass should
not vary much if the molecular cloud heating rate from starlight and
cosmic rays scales with the local stellar column density in the disk.
This is because molecular cloud cooling scales roughly with $T^2$ to
$T^3$ (Neufeld, Lepp, \& Melnick 1995), so the numerator in the
expression $T^2/P^{1/2}$ varies almost directly with the local gas
surface density of stars, and because the interstellar pressure scales
roughly with the square of the disk column density, so the denominator
in this expression also varies with the local stellar surface density.
Only when the star formation rate is very high per unit gas mass, as in
starburst regions, will the heating luminosity per unit gas mass
increase substantially, thereby increasing $T^2$ more than $P^{1/2}$ to
shift the ``characteristic'' or minimum mass of star formation upward.
This increase in star formation rate per unit gas mass is expected at
high average gas density, because the star formation rate scales with a
power of the gas density that is greater than 1 (Kennicutt 1998). Thus
the peak in the IMF should shift toward higher mass in star-forming
regions with very high ambient gas density. Note that this implies that
{\it the global efficiency of star formation should increase directly
with the minimum stellar mass}. Similarly, the global gas consumption
time should vary with the inverse of the minimum stellar mass. Other
than this physical variation of the low mass flattening limit with
$T^2/P^{1/2}$, the IMF should be ``universal'' according to this
theory, with statistically reasonable fluctuations that depend on the
number of stars sampled.

The model requires very few physical assumptions, yet many of the
physical processes that lead to star formation should be important for
the final stellar mass. Can the IMF really be so independent of the
processes of star formation? Of course, star formation is not purely
random as in the model: stars form when and where they do for physical
reasons. Indeed, the more we know about star-formation, the more we will
be able to say that a particular star formed for a particular reason. We
can say this for many stars already: there are known triggering
processes that form stars in well observed regions, either one star at a
time (as in pressurized globules) or in groups (as at the periphery of
HII regions; see the review of triggering processes in Elmegreen 1998).
These observations make random sampling models unnecessary. However, the
model uses random sampling only as a mathematical tool: there is an
underlying assumption that the mixture of real triggering processes is
about the same everywhere once more than several hundred stars are
formed. This means that the same fraction of stars forms by globule
squeezing, or clump collisions, or gravitational collapse along
filaments, or whatever the process might be, in all regions {\it on
average}. If this is the case, then we do not have to say how a
particular star formed in order to explain the IMF. By the time all of
these star formation processes are {\it reduced} to the IMF, by gross
averaging, essentially all of the important physical details have been
lost. In this sense, the IMF model is not a theory of star formation,
nor does it contain a theory of star formation. It is only a
representation of how a large number of processes can combine to give an
average stellar mass distribution that is subject to only a few
reasonable physical constraints, such as the basic structure of clouds
and the expected lower mass limit from cloud self-gravity.

An important assumption of the model is that star-forming cores
generate, or preserve from some previous state, a hierarchical structure
that is similar to what is commonly observed in non-star-forming clouds,
e.g., as fractal structure along molecular and atomic cloud edges (Beech
1987; Bazell \& D\'esert 1988; Scalo 1990b; Dickman, Horvath, \&
Margulis 1990; Falgarone, Phillips, \& Walker 1991; Falgarone, Puget \&
Perault 1992; Zimmermann \& Stutzki 1992, 1993; Henriksen, 1991; Hetem
\& Lepine 1993; Vogelaar \& Wakker 1994; Pfenniger \& Combes 1994; Fleck
1996; Elmegreen \& Falgarone 1996; Falgarone \& Phillips 1997; Stutzki
et al 1998). Hierarchical structure is critical to the theory, yet it is
not directly confirmed by observations of cluster-forming cores. Indeed,
most stars form in clusters (Lada, Strom, \& Myers 1993), but most young
clusters are too far away to have their gas structure resolved at the
0.1 M$_\odot$ level (Lada et al. 1991; Phelps \& Lada 1997).
Nevertheless, there is indirect evidence for structure in
star-forming cores in the form of extinction fluctuations (Lada et al.
1994; Alves et al. 1998) and a high level of collisional excitation
(Falgarone, Phillips, \& Walker 1991; Lada, Evans \& Falgarone 1997). 

The observation of stars in truly structureless cores is not
contradictory either. The only thing the model assumes is that the mass
which eventually ends up in stars comes originally from the ``branches''
in a hierarchical tree of gas structure (see Houlahan \& Scalo 1992
for a tree-like representation of cloud structure).
{\it When} this selection of
pre-stellar masses actually occurs in the life of a cloud does not have
to be specified to get the IMF. Indeed, most observations of
star-forming cores, particularly those with only one or a few stars, are
made at a phase in their evolution that is far removed from the time
when they were part of any fractal structure. The cores have already
become strongly self-gravitating, rounded, centrally condensed, and
dynamically {\it detached} from the turbulent structure around them
(e.g. Myers 1985; Andre, Ward-Thompson, \& Motte 1996), i.e., they may
not be fractal anymore. To understand star formation fully, we have to know
where these cores came from and what they were like {\it before} they
became strongly self-gravitating. But to understand the IMF in this
model, we only have to assume they evolved at constant mass from one of
the nodes of a hierarchically structured, younger cloud. 

Other theories of the IMF and star formation differ in a significant way
from this. Most assume that stars form in uniform clouds, perhaps moving
and accreting gas uniformly (e.g., Bonnell et al. 1997), or
spontaneously triggered by magnetic or other fluctuations in a uniform
or gradually stratified density (Carlberg \& Pudritz 1990; Myers 1998).
Others assume stars determine their own mass by wind-limited accretion
from a uniform medium (Nakano, Hasegawa, \& Norman 1995; Adams \&
Fatuzzo 1996) or by wind-dominated cloud dynamics (Silk 1995). The
observation of pervasive clumpy structure with a hierarchical design in
pre-stellar clouds (see reviews in Scalo 1985, Falgarone 1989, Elmegreen
\& Efremov 1998) suggests that these uniform cloud models may have the
evolution backwards. Star formation, we assert, is a progression from
hierarchical, non-self-gravitating clouds that probably form by
turbulent processes at modest pressure in the general interstellar
medium, to stellar-mass, centrally-condensed cores that are dominated by
gravity. Since the hierarchical structure in pre-stellar clouds goes far
below stellar masses (e.g., Heithausen et al. 1998), this evolution
necessarily involves {\it increased smoothing with time} at the stellar
mass level. The other models take the opposite approach by trying to
generate {\it increased substructure with time} inside initially smooth
clouds. 

Our approach, based on increased smoothing in clouds that are already
highly structured, may involve several physical processes. One might be
the coalescence of non-self-gravitating, sub-stellar clumps into
self-gravitating stellar clumps (Whitworth et al. 1998). This
preserves hierarchical structure if every clump coalesces with another
clump, and it preserves the density dependence for the conversion rate
of gas into stars because the collision rate of sub-structures is
proportional to the clump crossing rate on all scales (cf. Appendix).
Another process might be the progressive loss of magnetic wave coupling
to turbulent motions during cloud contraction, as the opacity to
ionizing radiation increases and the coupling of charged grains to the
magnetic field decreases (Myers \& Khersonsky 1995). Perhaps some of the
details of the IMF will depend on this evolution, but as long as the
probability that a node in the initial hierarchy turns into a star is
about the same as that assumed here, namely that it increases slightly
with local density, then a Salpeter IMF should result. 

\section{The Model: Variations in the slope of the IMF power law}

\begin{figure}
\vspace{5.5in}
\includegraphics{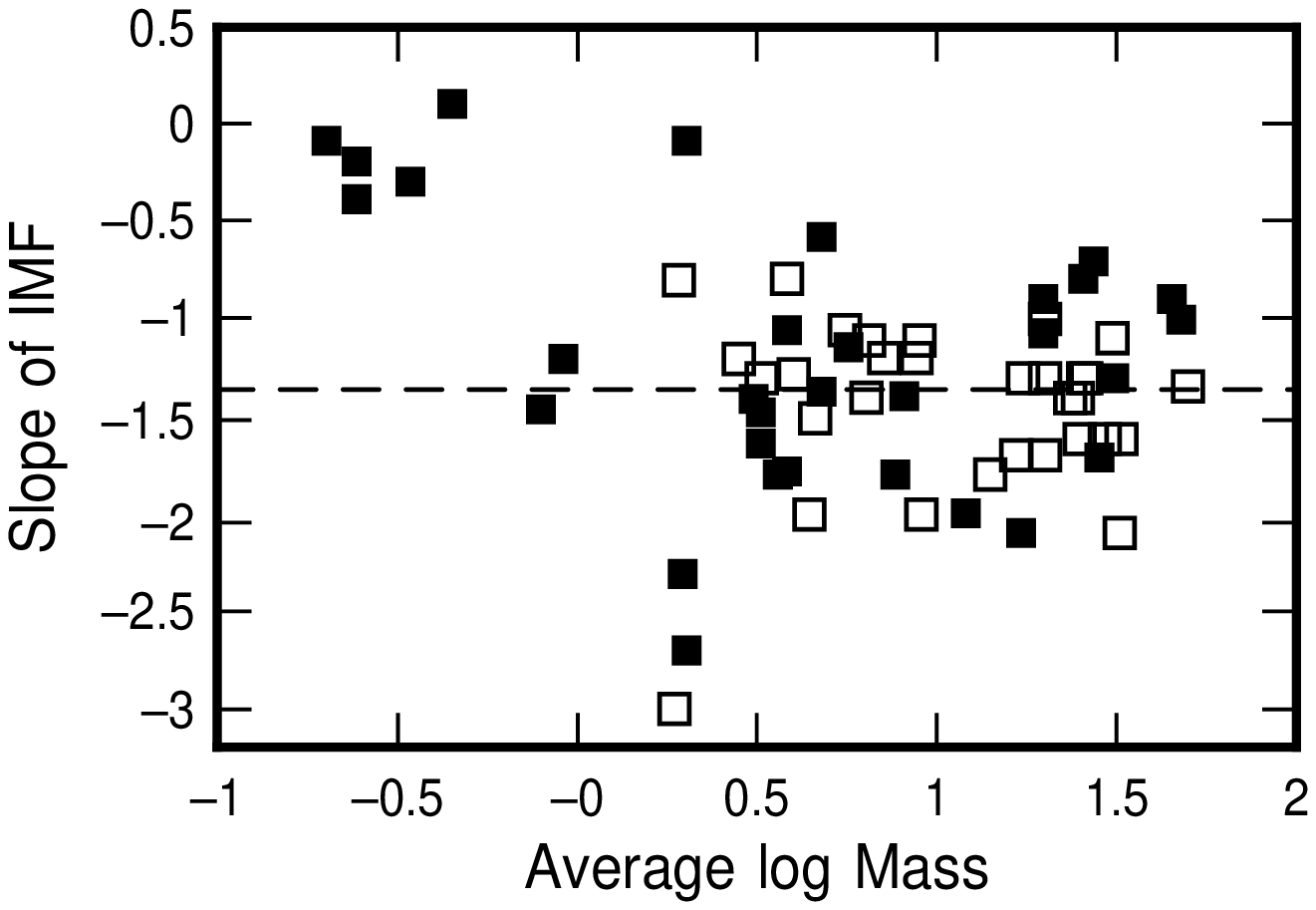}
\caption{IMF slopes in different clusters as a function
of the average log mass, in M$_\odot$, from Scalo (1998). 
The Salpeter value of $-1.35$ is shown by a dashed line.  Solid squares
are for clusters in the Milky Way, and open squares 
are for the Large Magellanic Cloud.}
\label{fig:scalo}
\end{figure}

Scalo (1998) compiled observations of the IMF and plotted the slopes
derived by various authors versus the average log mass of the
stellar sample. Scalo's diagram is reproduced in Figure \ref{fig:scalo}.
The result shows the general change in slope from a value close to 0 in
the flat part to a value that hovers around the Salpeter slope of
$-1.35$ (dashed line). These values are for the mass function when it is
plotted in equal intervals of the log of the mass; for equal intervals
of the mass itself, the slope is about $-1$ at low mass, and $-2.35$ at
intermediate to high mass. The figure shows the slope of the IMF varies
from region to region by about $\pm0.5$. There are typically several
hundred stars in each IMF determination (Scalo, private communication).

The average mass range on the abscissa for the samples in figure
\ref{fig:scalo} varies from sample to sample because the total number of
stars and the sensitivity of the surveys vary. Generally, the IMF is
determined from the most massive stars in a region because these are the
only stars that can be seen. If a region has only several hundred stars
total, then the most massive star is likely to be small and the average
mass plotted in figure \ref{fig:scalo} is small. If the region has
$10^4$ stars or more, but is far away and only the brightest stars can
been seen, then the average mass plotted in figure \ref{fig:scalo} is
large. The total number of stars in each survey is related to the mass
range for the fit. In the power law part of the IMF, with a power
$x=1.35$ as in the Salpeter IMF, the mass function in equal intervals of
mass is $n(M)dM =n_0M^{-1-x}dM$. If the largest star in the region has a
mass $M_L$, then $\int_{M_L}^\infty n(M)dM=1$, giving $n_0=xM_L^x$.
Suppose now that the mass range for stars in the fit to the slope is a
factor of $F$. Then the number of stars used for the fit is
$\int_{M_L/F}^{M_L} n(M)dM=F^x-1$. For $F=50$ and $x=1.35$, there are
$50^{1.35}-1\sim200$ stars in the sample.  Many observational surveys
have much smaller $F$ and number of stars than this. 

We fit the IMFs in the models in the same way they were fit in the
observations, using only the most massive stars in the sample. The model
was run 100 times, with a different number of stars each time, so the
most massive star in the sample varies from model to model. We chose a
mass range for the slope-fitting region to have its high mass limit less
than the mass of the most massive star in the sample by a fixed number
of filled histogram bins, equal to 20 bins in the cases shown here,
which corresponds to a factor of $\sim3$ in mass (the bin spacing is
logarithmic in mass). This avoidance of the most massive stars in the
ensemble puts us in the power law range, away from obvious statistical
fluctuations with a small number of most massive stars. It is what an
observer would do with an IMF that has a few ``odd'' massive stars at
the tail end of a smooth power law distribution. Then we systematically
decrease the lower mass limit
for the fitted region until a fixed total number
of stars is reached, which is chosen to be 200 to place us in the range
of typical IMF observations. This number, as well as the bin spacing,
will be varied to show how the rms fluctuations in the slope of the IMF
depend on the number of stars in the sample.

\begin{figure}
\vspace{5.5in}
\includegraphics{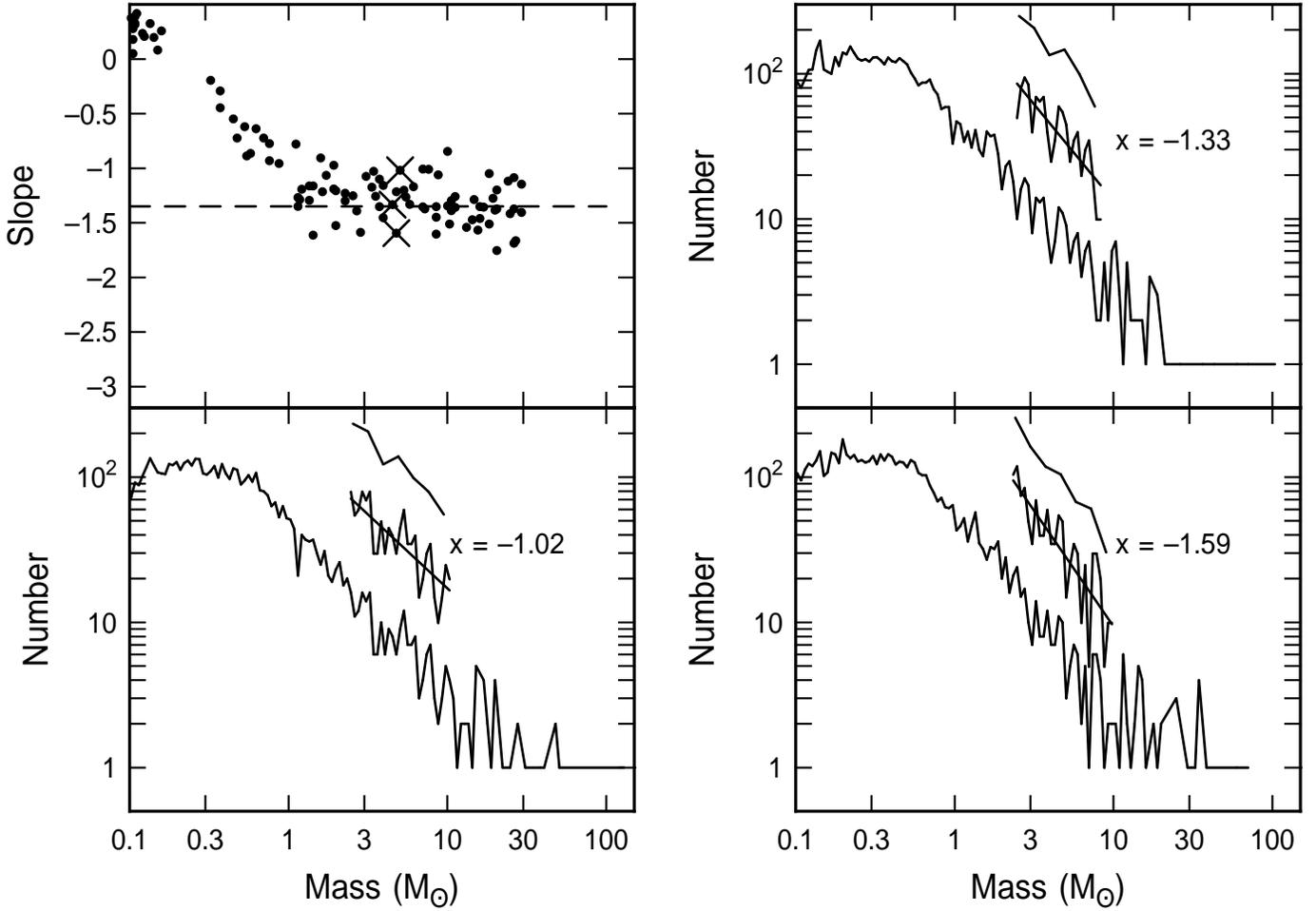}
\caption{(top left) IMF slopes in 100 models that differ in the
sequence of random numbers used to generate and sample the fractal
cloud tree, plotted as a function of the average
logarithm of the mass, as in Fig. 1. Each IMF slope is fit using
200 stars, with the largest star in the fit taken to be smaller 
by 20 filled mass bins than the largest star produced in the model.
The Salpeter value of $-1.35$ is shown by a dashed line.  The three
values indicated by crosses have their complete IMFs shown in
the other panels. The fitted portions of these IMFs are indicated
by the offsets, with two different bin sizes shown.} 
\label{fig:fluctuation}
\end{figure}

The top left panel in figure \ref{fig:fluctuation} shows IMF slopes for
100 random models, each with a different number of stars. The models all
have 8 hierarchical levels and an average of 3 sub-clumps per clump
(distributed as a Poisson variable -- see Paper I). The slopes were all
calculated at the high mass ends of the resulting IMF power laws,
including 200 stars when there were that many (the lowest mass points
have fewer than 200 stars), and excluding the stars in the 20 largest
mass bins for each sample, as discussed above. There is considerable
scatter at intermediate to high mass in the slope of the IMF around the
average value of $-1.35$, which is indicated by the dashed line. The
average slope increases to zero at low mass, as in the observations,
because this is where the IMF flattens out at the thermal Jeans mass
(assumed to equal 0.3 M$_\odot$ in the model). The slope even increases
to slightly positive values at large mass
in the model, because a Gaussian probability
for failure ($P_f$ in the notation of paper I) is used, and this causes
the IMF to fall at masses lower than the thermal Jeans mass. This
positive excursion of the IMF is not a physical result, but is based
entirely on this assumption of a Gaussian. Models with an exponential
$P_f$ had slopes increase only to zero, with no positive values. There
are no observations of the IMF slope at such low masses anyway, so we
have no particular reason to choose one $P_f$ over any other.

The other panels in figure \ref{fig:fluctuation} show the model
IMFs for the three cases indicated by crosses in the top left plot.
These are three representative cases illustrating how the IMF varies
around the mean slope. A case with the mean slope is in the top right,
and cases with shallower and steeper slopes are shown in the lower left
and right. The total model IMF is shown for each case, along with the
subpart that was used to calculate the slope for the 200 bright stars.
These subparts are shifted upwards for clarity, once with the same bin size
and again with the bins four times wider. 
The least square fits used for the plotted slopes are shown.

Observations of IMFs like those in the lower left and right of figure 
\ref{fig:fluctuation} might lead one to conclude that the power
law in the IMF varies significantly from region to region. 
However, these are just random
fluctuations in the model, both in the geometrical arrangement of the cloud
hierarchy, and in the selection of pieces to make stars. 

\begin{figure}
\vspace{5.5in}
\includegraphics{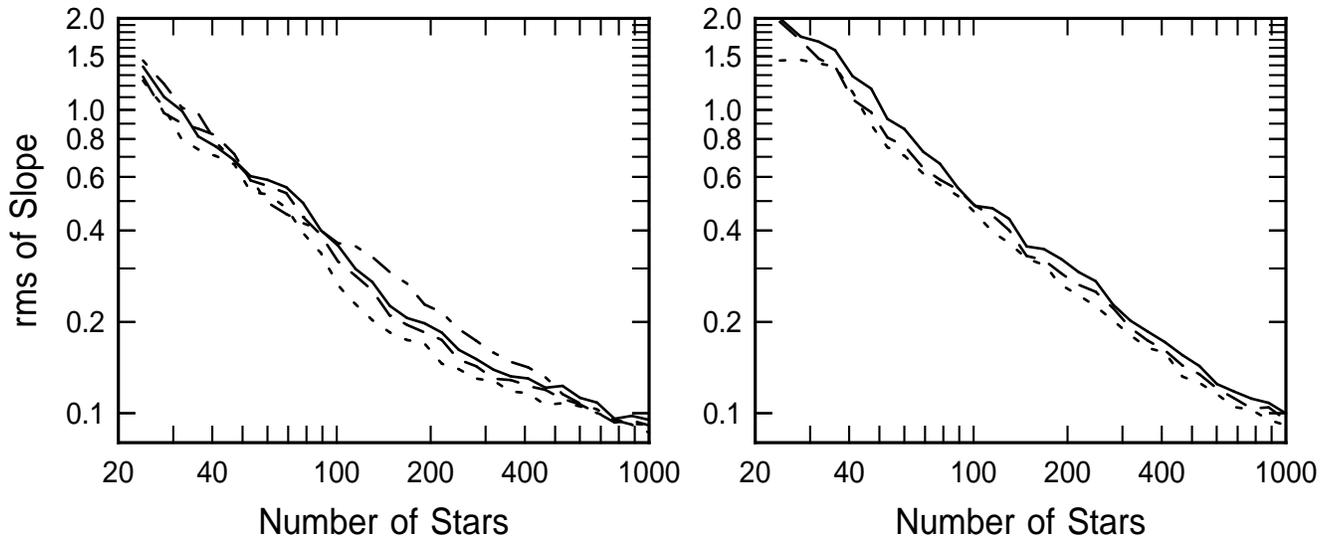}
\caption{(left) The rms deviation in the fitted slopes of the model IMFs for
the 1-10 M$_\odot$ range, plotted versus the number of stars
that are included in the fit.  The solid line is for
the case shown in Fig. 2, the dashed line has twice the bin
size, and the dotted line has four times the bin size.  The 
dot-dashed line is for another
100 independent IMFs that have an additional randomness in the
ratio of the star mass to the clump mass.  This figure shows that
the model IMF converges to a universal power law as the number of stars
in the cluster increases. (right) The rms deviations are shown for three
bin sizes in a model where a power-law IMF is sampled randomly, without any
consideration of how the power law arises. }
\label{fig:changes}
\end{figure}

Figure \ref{fig:changes} (left) shows the rms deviations of the model IMF
slopes from the average values in the mass range from 1 to 10 M$_\odot$
as functions of the number of stars that were used
to determine the slopes from each sample. The total numbers of stars in
the model clusters were equal to or larger than 1000 for this mass
range, ranging from $10^3$ to $1.8\times10^4$ as average stellar mass
increases from 1 M$_\odot$ to 10 M$_\odot$.
Three different bin spacings for logarithmic mass intervals are represented
by three different line types, corresponding to
factors of $3^{1/2}$, 3, and $3^2$ in mass for 10 bins. The figure
indicates that the rms fluctuations decrease with increasing number of
stars in the measurement, independent of bin size; this
is a reasonable result for statistical
fluctuations.  The mean number of stars per IMF fit that is required by
the model to give the observed level of fluctuations in real star
clusters, namely, $\pm0.5$, is about 80,
provided all the scatter in the observed slopes is from statistical
sampling. If some of the observed scatter is from measurement errors, which is
likely, then the mean 
number of stars necessary to reproduce the observed scatter 
could be larger.

The dot-dashed line on the left in figure \ref{fig:changes} represents a completely
different set of 100 models. These also produce scatter
plots like those in figure \ref{fig:fluctuation},
with the same average IMF slope, but there is an
additional statistical variation for this dashed line. 
This is an addition step in the model that
allows the fraction of the clump mass (from the hierarchical tree) that
actually goes into a star to vary in a random way. We use a randomly
generated mass fraction per clump, $f$, that is given by
\begin{equation} f={{e^{\xi}}\over{1+e^\xi}}\end{equation} where $\xi$
is a random variable distributed as a Gaussian centered on $\xi=0$ with
a dispersion of unity. This fraction $f$ therefore varies between 0 and
1 with a most likely value of 0.5, where $\xi=0$, and $f$ has a
variation around this most likely value ranging from an average of
$e^{-1}/(1+e^{-1})=0.27$ to $e/(1+e)=0.73$, representing a factor of 2.7
in the local efficiency of star formation. Such an additional random
variation seems reasonable for real star formation, but it does not
increase the rms variation in the IMF slope in any noticeable way,
according to figure \ref{fig:changes}.

The three lines on the right in figure \ref{fig:changes} were generated for
the same three bin sizes by randomly sampling
points directly from a power law with a slope $-(1+x)=-2.35$.  
There is no hierarchical
cloud structure or selection of cloud pieces as in the IMF model discussed
elsewhere in this paper.  The point of this diagram is to show the level of
statistical fluctuations that are expected from any model of the IMF that produces
an average power law with the same slope as in the hierarchical
model.  As in the other cases, we
first generate the stars, and then we select the power law region of the
sampled IMF starting a factor of 3 below the largest stellar mass. Obviously, 
random samples from any model of the IMF will produce fluctuations in the slope
when the number of stars is small.  It is interesting that these fluctuations
are slightly larger for the initial power law than for the IMF that was
generated by randomly sampling a hierarchical cloud. 

\begin{figure}
\vspace{5.5in}
\includegraphics{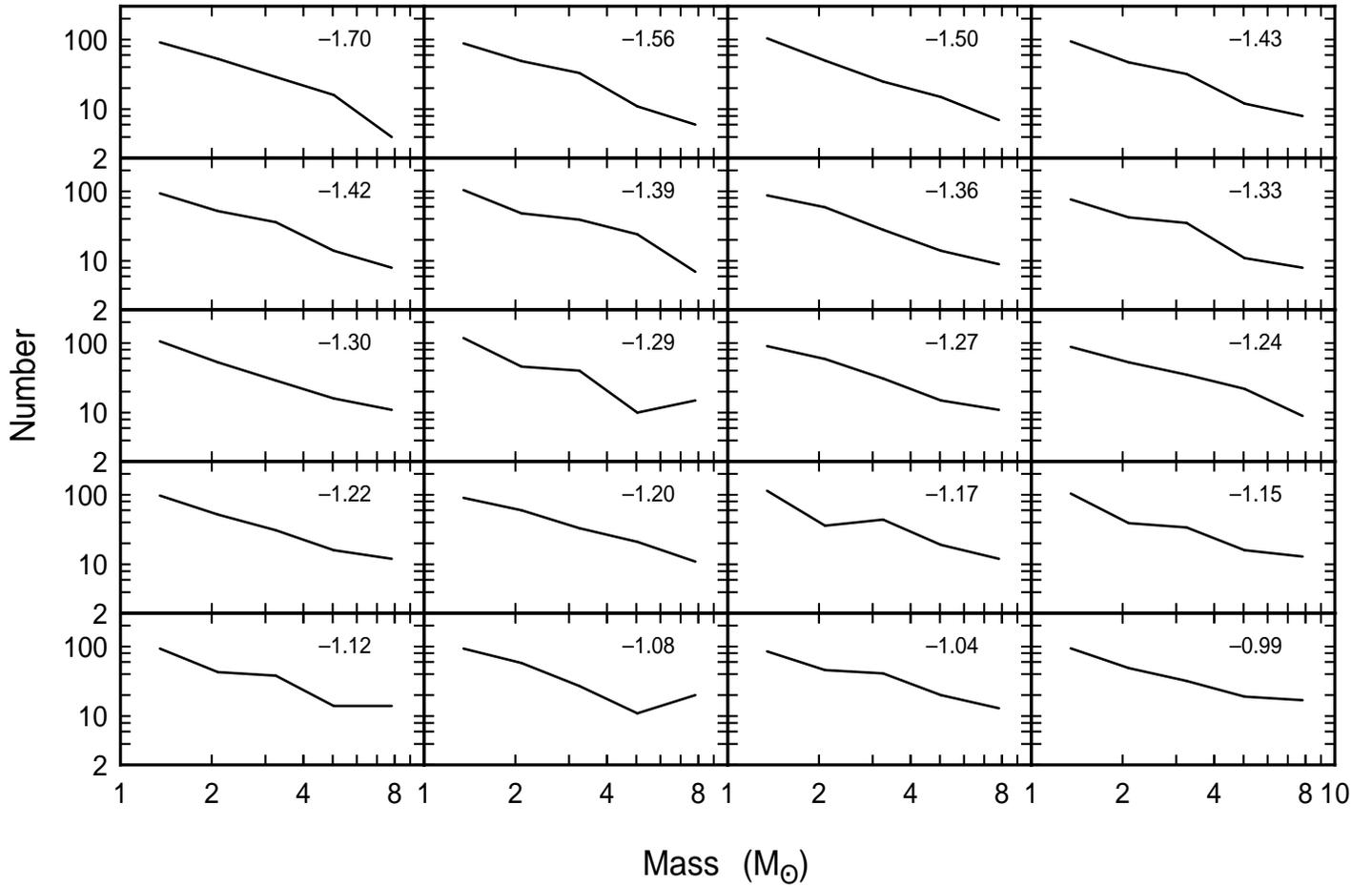}
\caption{(left) The IMFs from 20 out of 100 runs, selected 
in a regular and unbiased way to represent the  
full range of slopes, along with values of the slopes in each
panel, evaluated in the mass range from 1 to 10 M$_\odot$. }
\label{fig:lots}
\end{figure}

\begin{figure}
\vspace{5.5in}
\includegraphics{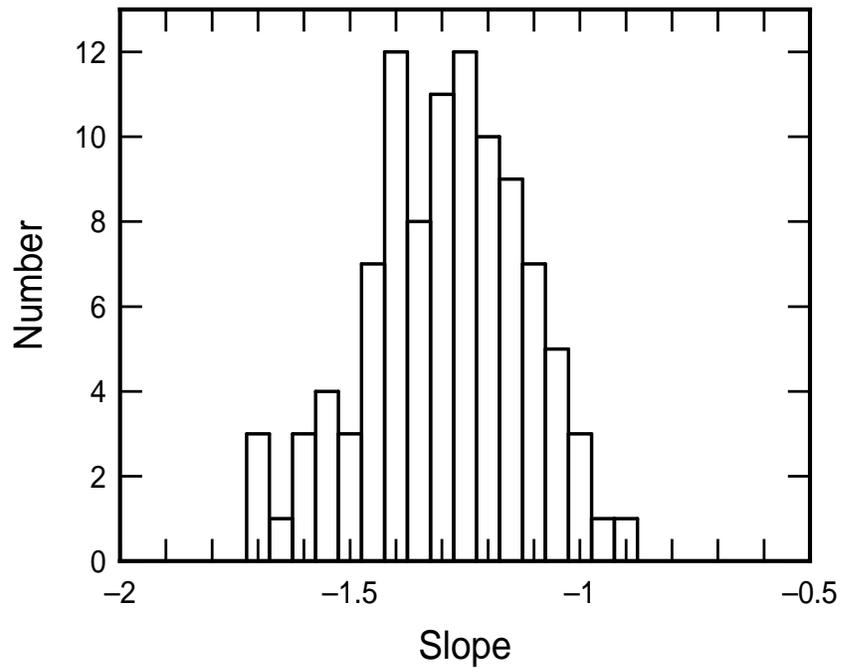}
\caption{(left) Histogram of 100 slopes in the mass range 
from 1 to 10 M$_\odot$ for 100 standard IMF models.}
\label{fig:hist}
\end{figure}

Figure \ref{fig:lots} shows 20 model IMFs in the range from 1 to 10 M$_\odot$,
selected out of a new sample of 100 random IMFs 
constructed from hierarchical trees in the usual way. 
These 100 models were all chosen to have
a total number of stars ($\sim2400$) such that 
$\sim200$ were in the 1--10 M$_\odot$ mass range.
To make the figure, 
the power law slopes for the resultant IMFs were evaluated
in this mass range, and placed in ascending order. Then 
every fifth IMF was chosen to get the 20 plotted IMFs .
This gives an unbiased choice for presentation that spans the full range
of IMFs from these 100 runs. The 
values of the slopes are shown in each panel.
There are 5 bins between 1 and 10 M$_\odot$, to be consistent
with the small number usually used by observers; this choice
of bins should not affect the slopes (cf. Fig. \ref{fig:changes}).
A histogram of all 100 slopes for this sample, again evaluated
in the 1--10 M$_\odot$ mass range, is in figure \ref{fig:hist}. 
The dispersion in this histogram 
is consistent with the value obtained from figure \ref{fig:changes}
for the cases with 200 stars in the 1--10 M$_\odot$ mass interval. 

The IMF fluctuations shown in this section represent what should be
expected from random variations around a universal IMF. There is no
a priori reason to think that the real IMF is universal, but if it is, 
then slope variations of several tenths, depending on the sample size,
are to be expected.  If the real IMF is not universal, but has
variations in power law slopes that are greater than 
statistical fluctuations,
then additional physics has to be added to 
the model, or the model is wrong.  For example, Scalo (1998) gives
examples of well-studied clusters with smooth power law IMFs that
vary in slope from $x=1.1$ to 2.3 (NGC 663 from Phelps \& Janes 1993; 
Pleiades from Meusinger et al. 1996). If these observations reflect the 
real initial mass functions for the entire clusters, including
the outlying stars, and not just the effects of mass
segregation, sampling errors, improper evolution corrections,
and other problems, then the theoretical model needs revision.   

\section{The Model: Mass-dependent Birth positions}

The IMF model in Paper I operates by first generating a random hierarchical
tree including a wide range of masses at the various branches, ranging
from masses much less than a stellar mass to masses generally larger
than the largest star. It then selects branches from this tree randomly
and assigns the residual mass on that branch to a new stellar mass that
becomes part of the final IMF. This selection of masses is repeated for
each star until the original tree has no mass left (or has some fraction
of its initial mass left, in an alternative model -- see Paper I), and
then a new tree is generated and new star selections begin. 

The generation of the hierarchical tree was described in detail in Paper I.
Now we add positions to this tree, so that each node has a
three-dimensional coordinate that is consistent with the fractal
dimension of the overall structure. This dimension is assumed to be 2.3,
because this is the value we found from a survey of molecular cloud size
distributions (Elmegreen \& Falgarone 1996), but other values close to
this give about the same IMF. The relative positions of the ``inner''
subclumps inside each ``outer'' clump are determined by the addition of
random offsets to each outer clump position in three dimensions. This
offset depends on the level $h=0$ to 8 in the hierarchy, scaling as
$L^{-h}$ for scale reduction factor $L=1.6123$, which gives a fractal
dimension $D=\log 3/\log L=2.3$; here, the 3 in the numerator is the
average number of subclumps per clump in the model (other values were
also used for models in Paper I, with no significant differences in the
resulting IMF).

The spatial distribution of mass for the stars that form in the model is
determined entirely from this hierarchical tree structure. As for a real
tree, the trunk and massive branches are closer to the center of the
distribution of all branches than are the low mass branches. Thus the
high mass stars tend to form near the center of the cluster in this
model. This is also the sense of the birthplace distribution from
observations (Hillenbrand \& Hartmann 1998).

\begin{figure}
\vspace{5.5in}
\includegraphics{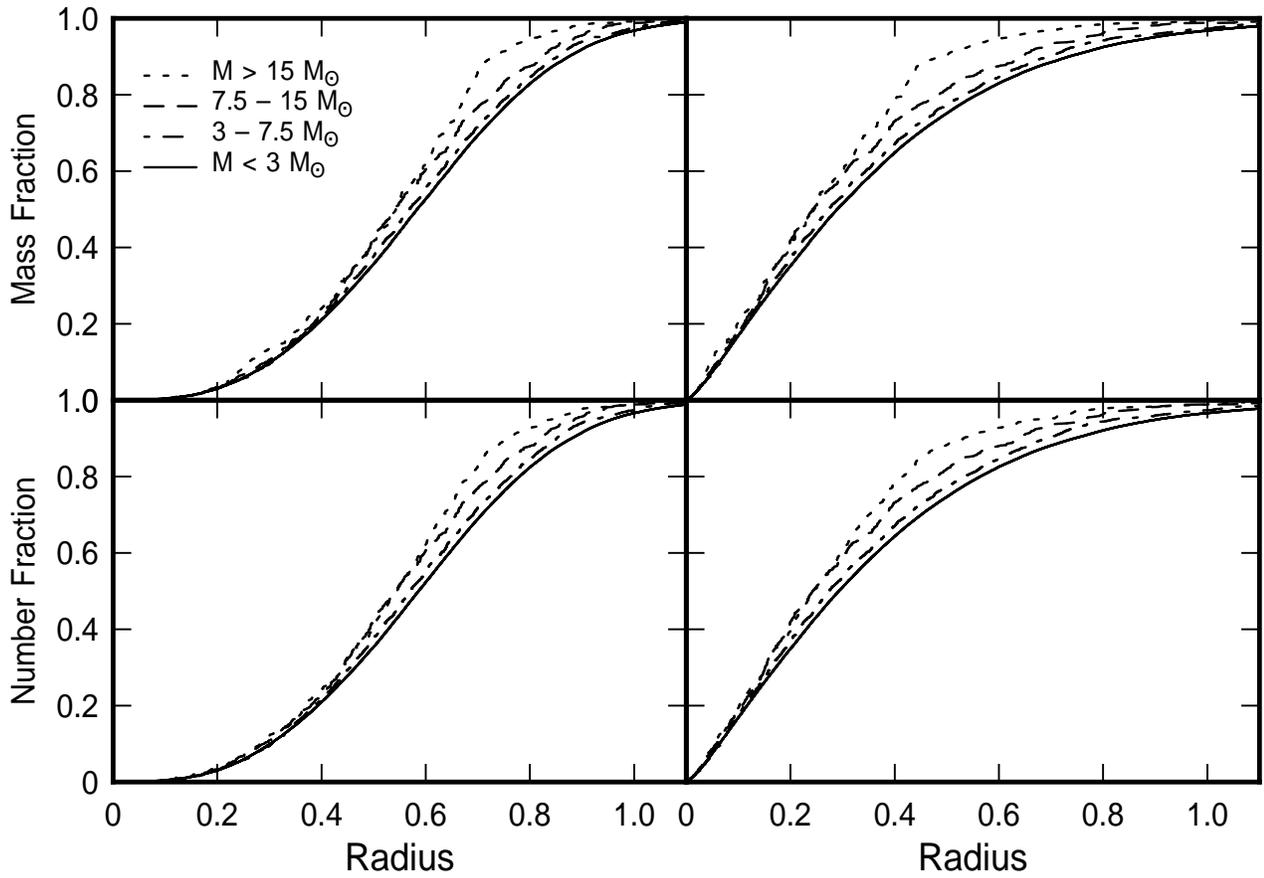}
\caption{The average birthplaces of stars with various masses are shown
as cumulative functions of the number fraction and mass fraction
within a certain radius.  The panels on the left are for the pure
fractal model, and the panels on the right are for a distorted
fractal with a density profile like that of a molecular core. The massive
stars are born slightly closer to the center than the low mass stars.}
\label{fig:birthplace}
\end{figure}

Figure \ref{fig:birthplace} shows the number fraction (bottom) and the
mass fraction (top) of stars having radii less than the value on the
abscissa, for a single IMF model with $10^5$ stars. This is the type of
plot used by Hillenbrand \& Hartmann to demonstrate a mass-dependent
birth position in Orion. In the left-hand panels, the increase in the
number fraction or mass fraction with radius traces out the density
structure in the model cloud. The four lines in each plot represent
different mass intervals for the stars, as indicated by the figure
legend. The lines for the most massive stars rise faster at small radii,
indicating that these massive stars are more centrally condensed than
the low mass stars. The effect is very weak, however.

The panels on the right in figure \ref{fig:birthplace} show the same
data as the panels on the left, but they are plotted with a distorted
coordinate system in which the radius $R^\prime$ equals the $2.3$ power
of the undistorted radius $R$. This is the type of transformation that
would be made by a hierarchical cloud as it becomes centrally condensed
by self-gravity to the commonly observed $R^{-2}$ density distribution
of an isothermal sphere. During such a transformation, which presumably
occurs before star formation begins, the gas mass is conserved so that
\begin{equation} \int_0^{R^\prime} n^\prime(r)4\pi r^2dr=\int_0^R n(r)
4\pi r^2 dr \end{equation} gives $R^\prime(R)$. Before the
transformation, $n(r)\propto r^{D-3}$ around the trunk of the fractal
tree (because mass scales with $R^D$). If we set $n^\prime\propto
r^{-2}$, then $R^\prime\propto R^D$. 

The model we have in mind here is one in which an interstellar cloud
begins its life in a non-self-gravitating state, dominated by turbulence
and other processes that give it an internal hierarchical structure down
to very small scales. Over time, this cloud contracts in the center as a
result of self-gravity and the smallest (sub-stellar) clumps begin to
merge to form a few star-forming cores. These cores are assumed to
preserve the mass distribution they had from the original hierarchical
structure, because they are still part of that structure going to larger
scales and because the cloud may continuously regenerate new
hierarchical structures and new cores as a result of continued
turbulence and other processes, including self-gravity. We get the same
IMF as in the original model because the overall cloud topology has not
changed, but we get stellar birth locations that are more condensed,
reflecting the final state of the gas at the time of star formation.
This model is purely phenomenological, but it seems reasonable given the
similarities and differences in the structures of diffuse and
self-gravitating clouds. 

The panels on the right of figure \ref{fig:birthplace} show the
distribution of birth positions in the condensed cloud. All the stars
are more centrally concentrated than they were in the uncondensed cloud,
and the concentration of the most massive stars is more pronounced. How
well this agrees with the observations is not clear because Orion is the
only case where this effect has been measured, independent of
evolutionary effects, and Orion has only a few massive stars.
Nevertheless, there is some tendency in hierarchical clouds to have the
massive stars born closer to the center than the low mass stars. A study
of birth positions similar to that for Orion but for 30 Dor would be
more revealing because 30 Dor has more stars. 

It may be that real star-forming clouds have a much stronger central
concentration of massive stars than than the centrally-condensed cloud
model predicts. Then additional physical processes must be involved.
This seems likely because the center of a cloud differs from the edge
with respect to the density, gravitational potential, and degree of
shielding from outside radiation. For example, the protostars near the
center could accrete gas at a higher rate and become more massive
(Larson 1978, 1982; Zinnecker 1982; Bonnell et al. 1997); they could
coalesce more (Larson 1990; Zinnecker et al. 1993; Stahler, Palla, \& Ho
1998; Bonnell, Bate, \& Zinnecker 1998), or the most massive stars and
clumps formed anywhere in the cloud could fall to the center faster
because of gas drag (Larson 1990, 1991; Gorti \& Bhatt 1995, 1996).

\section{The Model: Mass-dependent birth order}

The birth time of the {\it first} star that forms in a region with a
mass in a logarithmic interval centered around $M$ should increase as
$M^{1.35}$ if the star formation rate is constant and the IMF is
randomly sampled. This is true for any model of the IMF that is based on
a random appearance of stars of various masses. The {\it average} birth
times for stars should be independent of mass in these models, however.
The mass dependence for the time of {\it first} birth arises because the
number of stars in a logarithmic interval around $M$ is proportional to
$M^{-1.35}$ in our model and for a Salpeter IMF. Thus the mean time
between birth events for stars of this mass equals the total elapsed
time of star formation divided by the number of such stars, and this
gives a mean time interval between births $\propto M^{1.35}$. The mean
delay time before the first birth of a star of mass $M$ equals the mean
time interval between all births of such stars, for a uniform star
formation rate. Thus the mean delay time before the first birth of a
star in a logarithmic interval around $M$ increases with $M^{1.35}$. As
a result, an observer of a real star cluster will notice that the low
mass stars generally form before the first intermediate mass stars
appear, and these intermediate mass stars appear before the first high
mass stars. Yet, for a large enough sample of clusters, the {\it
average} time of appearance of each star is always about half the
current age of the region (for a constant and continuing star formation
rate).

This statistical effect is strong and should be obvious in an active
region. It is analogous to the well-known effect in which larger regions
tend to produce larger, most-massive 
stars (see section I). The timing result
follows from our model as well. We computed the IMFs for many clusters
and kept track of the time of formation of each star. Unlike the
previous models, we did not repopulate the cloud after it exhausted its
gas, but just quit when it did, with whatever stars formed. Typically
between 200 and 500 stars formed in each cluster with 8 levels in the
model hierarchy and 3 subclumps per clump. This is a reasonable size for
comparisons with observations. 

\begin{figure}
\vspace{5.5in}
\includegraphics{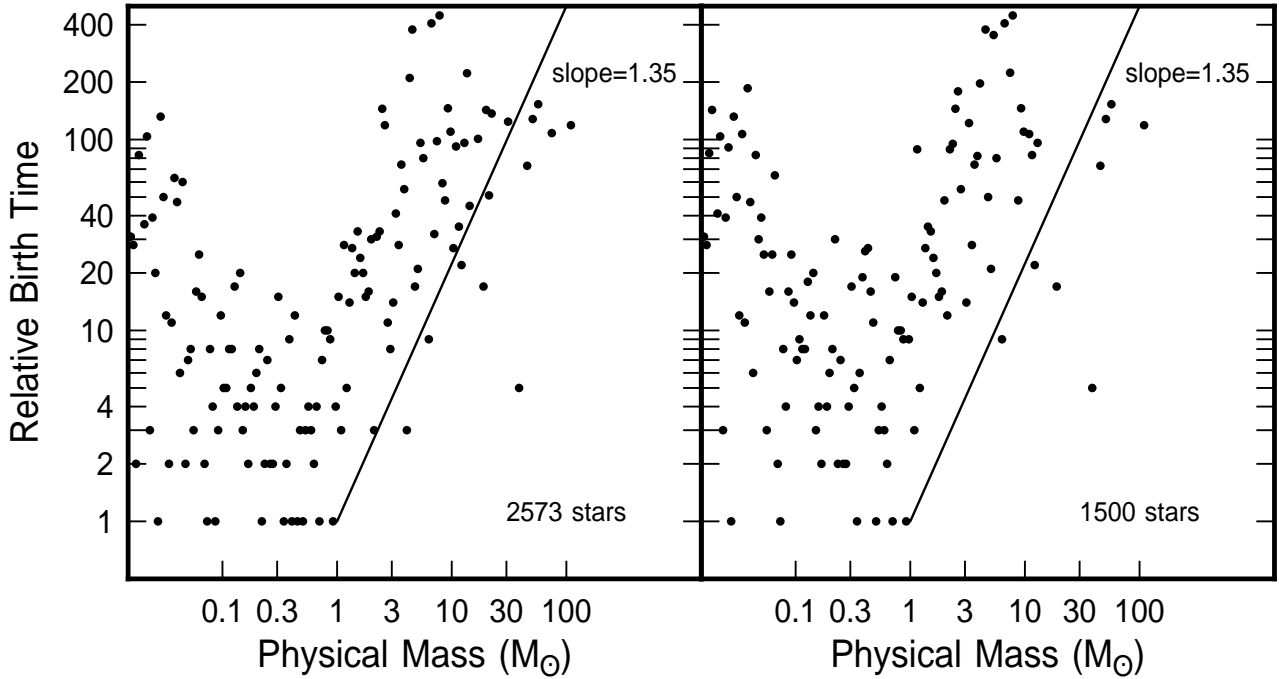}
\caption{The relative birth time for the first appearance of
a star with a certain mass is shown as a function of that
mass for a composite of 11 clusters on the left, containing
2573 stars total, and 6 clusters on the
right, containing 1500 stars. The relative birth time is measured
in units of the time of first appearance of any star in the model.
The dots with relative birth times of 1 are the first masses that
appear in the 11 or 6 clusters. The dots with large relative
birth times are for masses whose first appearance was late in 
every cluster. The solid line has a slope of 1.35, indicating the
theoretical prediction. Massive stars tend to form late in a cluster
because they are rare.  Thus the relative birth time has a distribution
with mass that is the inverse of the IMF.}
\label{fig:birthtime}
\end{figure}

Figure \ref{fig:birthtime} (left) shows the time of first appearance of
a star of a certain mass as a function of that mass, in logarithmic
intervals, for all of the stars that formed in 11 computed clusters.
Figure \ref{fig:birthtime} (right) shows the same result for 6 clusters.
In both cases, the distribution of points is the inverse of the IMF,
showing a minimum first time of appearance at the same mass where the
IMF peaks, and an increase in the time of appearance on both sides of
this minimum where the IMF decreases. Recall that these models use a
Gaussian probability of failing to form a star, $P_f$, and this is
entirely by assumption, so the decrease in the IMF and the increase in
the first time of formation at masses less than the peak in the IMF has
no observational basis and may not apply to real clouds. We could just
as easily have assumed an exponential $P_f$, as in some cases of Paper
I, and then the first time of formation would level off to a small value
at all low masses. The increase in the first time of formation for
masses higher than 0.3 M$_\odot$ is a prediction of the model,
reflecting entirely the stochastic nature of the distribution of star
birth in time. The precise form of this birth order distribution is not
known from observations, but we predict it to be proportional to
$M^{1.35}$.

\section{The Model: Other IMF cases}
\label{sect:alpha}

In Paper I and for all of the models discussed here so far, we have
assumed that the probability of selection of a clump in a hierarchical
cloud is proportional to the square root of the local clump density,
which is larger at smaller mass because of the nature of this type of
structure. As a result, low mass clumps are selected more often than the
proportion of their number, and the stellar mass spectrum is slightly
steeper than the clump mass spectrum. This is a sensible application of
various star formation theories, which suggest that gravitational
processes, magnetic diffusion, turbulent decay, and coalescence all
contribute in some fashion to the local rate of conversion of gas into
stars. Considering the molecular cloud scaling laws, all of these
processes operate on a timescale that scales approximately with the
inverse square root of the local density (see Appendix). 

Other models are possible. For example, the conversion rate from
non-self-gravitating clumps to self-gravitating clumps that eventually
make stars might be independent of density. We showed in Paper I that
this gives an IMF with too shallow a slope, $-(1+x)\sim-2$ instead of
$-2.35$, but we can tune the model by selecting only a fraction of the
chosen clump mass to go into the star. This is done by letting the star
mass, $M_s$, equal the chosen gas clump mass, $M_c$, raised to some
power, $1-\alpha$, so the {\it local} efficiency of star formation,
calculated on a star-by-star basis, is
\begin{equation}
{{M_s}\over{M_c}}\propto\left({{M_c}\over{M_J}}\right)^{-\alpha}
\end{equation}
for characteristic thermal Jeans mass, $M_J$, which is the mass at the
peak of the IMF. 

\begin{figure}
\vspace{5.5in}
\includegraphics{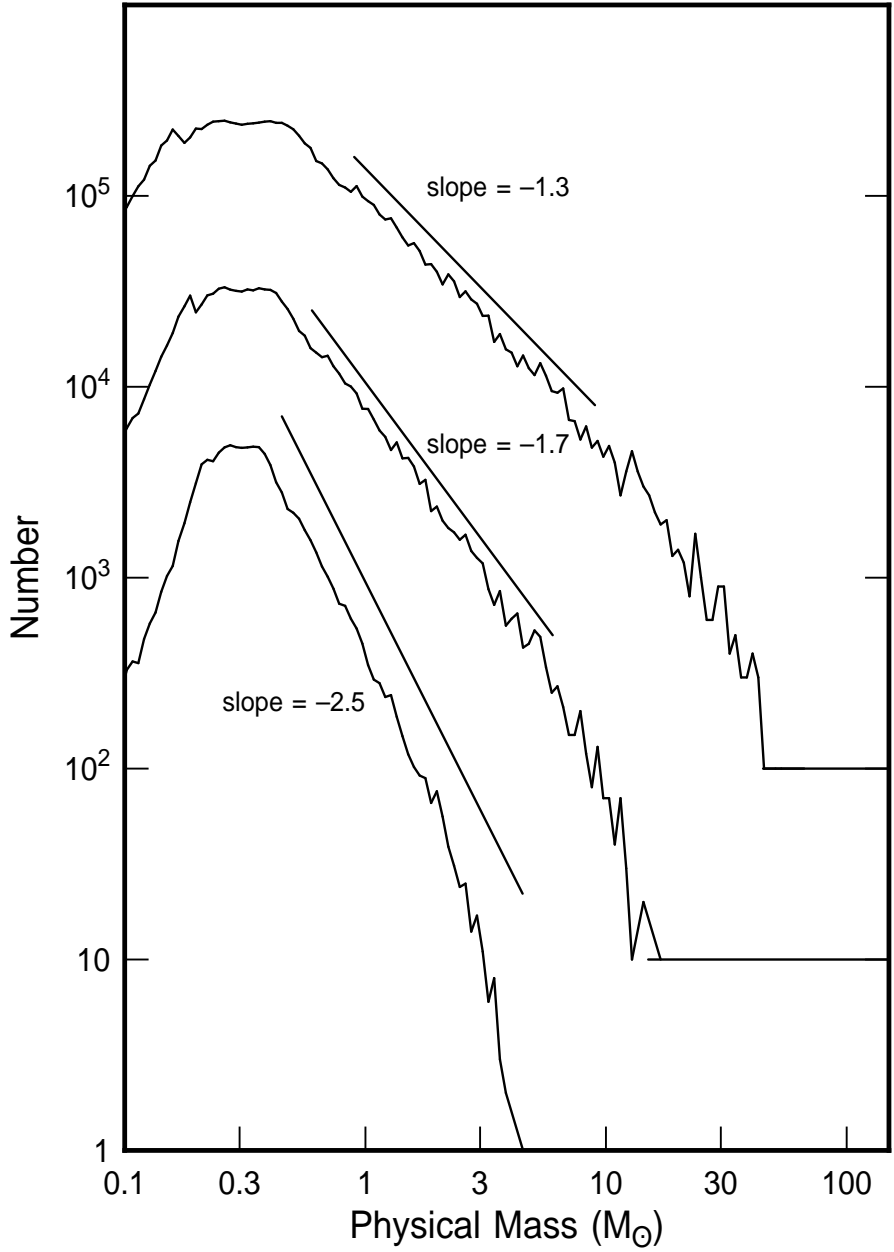}
\caption{Three model IMFs using $10^5$ stars 
are shown with different power law
distributions of the local efficiency, $M_s/M_c\propto M_c^{-\alpha}$,
for $\alpha=0.2$ on the top, 0.4 in the middle, and 0.6 at the
bottom. The solid lines show the predicted slopes. The bottom IMF
has the proper number given on the ordinate, while the top two 
IMFs are shifted upwards for clarity.}
\label{fig:3models}
\end{figure}

Figure \ref{fig:3models} shows the IMFs from three models with
$\alpha=0.2, 0.4,$ and 0.6; larger values have steeper slopes. The
actual slopes can be calculated by converting the protostellar cloud
distribution $n(M_c)d\log M_c\propto M_c^{-x}d\log M$ into a star
distribution $n(M_s)d\log M_s$ as follows:
\begin{equation}
n(M_s)=n(M_c){{d\log M_c}\over{d\log M_s}}\propto M_c^{-x}
= M_s^{{-x}\over{1-\alpha}}.
\end{equation}
For $\alpha=0.2, 0.4$ and 0.6, with $x=1$ in this case, the slope of the
IMF becomes $-1.25, -1.67$, and $-2.5$, in agreement with the results in
the figure. Thus the Salpeter IMF can be recovered even in the case
where there is a uniform selection probability per level, by choosing
$\alpha=0.3.$ Similarly, the steep IMF seen by Massey et al. (1995),
which has $x\sim4$, can come from our usual Salpeter model ($x=1.3$) if
$\alpha=2.7$. 

These solutions are not particularly enlightening because the
introduction of an additional parameter has no firm observational basis.
Theoretical work on a variable local efficiency has been considered
elsewhere (e.g., Zinnecker 1989; Nakano, Hasegawa \& Norman 1995; Adams
\& Fatuzzo 1996). We show the application of this concept here only to
emphasize that in our model, the underlying gas mass spectrum that has
to be modified by a variable local efficiency has $x\sim1.3$, which is
much steeper than the usual clump mass spectrum measured in molecular
cloud surveys, which have $x\sim 0.5-0.7$ for logarithmic intervals of
$M$. Thus any ``corrections'' that might be introduced to convert the
gas mass spectrum to a star mass spectrum are much smaller in our
model than in standard models. 

If the density dependence of the selection rate differs from both
powers 0 and 0.5, then the IMF can still be estimated from 
the basic assumptions of the model.  Suppose the selection rate 
scales with density as $\omega\propto\rho^\alpha$. The density
scales with mass as $M^{1-3/D}$, so the mass dependence of the
selection rate scales as $M^{\alpha\left(1-3/D\right)}$. This
mass function directly multiplies the IMF. For
$\alpha=0.5$ and $D=2.3$, as in the standard case considered for this
model, the selection rate $\propto M^{-0.15}$, so the IMF is
$M^{-2.15}$ without competition for mass, and it steepens to $M^{-2.3}$
with competition for mass. Evidently, larger $\alpha$ makes the IMF steeper. 

\section{Conclusions}

A model for the initial stellar mass function based on random selection
of gas pieces in a hierarchical cloud, with a selection probability
proportional to the square root of density and a lower mass cutoff from
the lack of self-gravity, has been shown here and in Paper I to
reproduce six features of the observed IMF: (1) the power law slope and
its remarkable constancy from place to place over the history of the
Universe, (2) the low mass flattening at about the thermal Jeans mass,
(3) the constancy of this thermal Jeans mass from place to place and
over time in normal star-formation environments, and its increase in
starburst regions, (4) the seemingly random fluctuations in the
power-law slope from region to region, (5) the tendency for higher mass
clusters and clouds to form higher mass stars, and (6) the tendency for
high mass stars to form more centrally condensed in clusters and at
later times than low mass stars.  The only physical scale involved in
any of these processes is the thermal Jean mass, which is about 0.3
M$_\odot$ in the solar neighborhood for typical temperatures and
pressures in star-forming clouds (Paper I).

Reasonable modifications to this theory make it more realistic in terms
of star formation processes, but should not affect the resulting IMF.
For example, star-forming clouds may evolve from a previous diffuse
state by simple contraction, or perhaps by high-pressure, forced
rearrangement (e.g., triggering), to form centralized concentrations
that have essentially the same hierarchical structure as the original
cloud, although the condensed version is not self-similar anymore
and the individual clumps may have redistributed their gas. 
Nevertheless, as long as the stars form in clumps that come from many
different levels in the hierarchy, the Salpeter IMF will result.

Another implication of the model is that the individual cores in which
single or binary stars form must have grown from lower mass clumps as a
result of various smoothing processes that counteract the tendency for
turbulence to subdivide. In this sense, our model differs significantly
from most other IMF models, which attempt to partition initially
uniform gas into stellar-mass pieces.  Our starting point is the
observation by several groups that hierarchical structure and power-law
mass distributions continue down to sub-stellar scales in
pre-star-formation clouds.

Additional observations and experimental tests of the model were
proposed.  Any direct observation of cluster-to-cluster variations in
the mass at which the IMF changes from the power law part to the flat
(or even turnover) part would give us a better idea of whether or not
this characteristic mass is the thermal Jeans mass, depending
exclusively on the square of the gas temperature and the inverse square
root of the cloud-core pressure.  More observations of the radial
distributions of various stellar masses in clusters are also necessary
to determine whether the physical processes of star formation change
significantly in the cloud core.  We would also like to know whether
deviations from a Salpeter IMF in clusters ever exceed the statistical
expectations from the theory, given the number of stars observed.

These and other tests are useful because they are based on observations
of stars rather than gas, so they represent a known state in the
evolution of the star-forming region. Observations of the gas have a
problem in not knowing this evolutionary state, e.g., whether a
relatively smooth, centrally condensed clump in which a star is
currently forming came from the fragmentation of an even larger, but
smooth, gas distribution, or the blending together of much smaller
structure that was made by turbulence. Observations of the gas also
suffer from resolution limitations and chemical selection effects. There
is also a tendency, when observing the gas, to overlook the wide variety
of processes that are likely to contribute to star formation,
concentrating instead on the boundary and other conditions of the
particular cloud under investigation. Such conditions are essential for
understanding the star formation process itself, but they may not be
necessary for understanding the IMF, which is a highly reduced average
of all possible outcomes. 

\section{Appendix}

The time scales for many processes connected with star formation are
roughly proportional to the inverse square root of the local density. This is
obviously true for self-gravitational processes, which operate on the
time scale $\left(G\rho\right)^{-1/2}$ for local density $\rho$. It is
also true for magnetic diffusion when the gas is in equipartition
between magnetic and self-gravitational forces, or between magnetic,
turbulent and self-gravitational forces, because then the diffusion
time is a constant of order 10 times the gravity time (for
detailed models of this, see Elmegreen 1979 or
Nakano 1998, and references therein; for a general review, see
Shu, Adams \& Lizano 1987).

Turbulent processes work on the same time scale. The turbulent crossing time
is $L/v$ for length scale $L$ and turbulent speed $v$, but the scaling
relationships for molecular clouds (Larson 1981; Solomon et al. 1987) give
$v\sim L^{1/2}$
and $\rho\propto L^{-1}$, so $L/v\propto\rho^{-1/2}$. For clouds in 
equipartition between magnetic, turbulent and self-gravitational forces, 
the constant of proportionality here is about the same as that for 
gravitational processes. 

Collisions between subclumps at each level also have a time scale of
about one crossing time on the scale of the larger clump in which they are
clustered. If the ``internal'' subclump sizes are $L_i$, the size of
the ``external'' clump that contains them is $L_e$, and the turbulent
speed on the external scale is $v_e$, then the collision time of the
subclumps inside each clump is $t_{col}\sim\left(n_i\pi L_i^2 v_e\right)^{-1}$
for subclump density $n_i=N_i/L_e^3$ and subclump number $N_i$.  
Substituting this expression for $n_i$, we get $t_{col}
\sim \left(N_i\pi L_i^2/L_e^2\right)^{-1}L_e/v_e$. 
For a fractal cloud, $N_i=\left(L_e/L_i\right)^D$ with fractal dimension $D$,
thus $t_{col}$ 
equals $\left(L_i/L_e\right)^{D-2}/\pi$ times the crossing time, $L_e/v_e$.
If the hierarchical structure is geometrically self-similar, then
$\left(L_i/L_e\right)^{D-2}$ is constant for each clump and equal to $N_i^{-(D-2)/D}
\sim N_i^{-0.13}$ for $D=2.3$.  Considering $N_i\sim3$ in our IMF models, this
is $0.87$. Thus the collision time of internal clumps 
is about equal to the crossing time of the external clump that contains them,
and this is true for all scales. 
This time is also approximately $(G\rho)^{-1/2}$ when the clumps are
virialized.

Helpful comments from John Scalo are greatly appreciated.

\newpage
\newpage
\newpage
\newpage
\newpage
\newpage
\newpage
\newpage
\end{document}